\documentstyle[12pt]{article}\textheight 235mm\textwidth 160mm
\newcommand{\bi}{\bibitem}
\newcommand{\be}{\begin{eqnarray}}
\newcommand{\ee}{\end{eqnarray}}
\newcommand{\nn}{\nonumber}
\begin{document}
\hspace*{11.6cm}KANAZAWA-93-01
\vspace*{2cm}
\begin{center}{\bf Unrecognizable Black Holes in Two Dimensions}
\end{center}
\vspace*{1cm}
\begin{center}{Takanori Fujiwara$\ ^{(1),(a)}$, Yuji Igarashi$\ ^{(2),(b)}$
and  Jisuke Kubo$\ ^{(3),(c)}$}
\end{center}
\vspace*{0.5cm}
\begin{center}
{\em $\ ^{(1)}$ Department of Physics, Ibaraki University, Mito 310, Japan}\\
{\em $\ ^{(2)}$Faculty of General Education, Niigata
University, Niigata 950-21, Japan}\\
{\em $\ ^{(3)}$ College of Liberal Arts, Kanazawa University, Kanazawa 920,
Japan }
\end{center}
\vspace*{1cm}
\begin{center}
ABSTRACT
\end{center}
The classical 2D cosmological model of Callan, Giddings, Harvey and
Strominger possesses a global symmetry that is responsible
for decoupling of matter fields. The model is
quantized on the basis of the extended phase space method to allow
an exhaustive, algebraic analysis to find potential anomalies.
Under a certain set of reasonable assumptions we show
that neither the BRST symmetry
of the theory nor
the global symmetry suffers from an anomaly. From
this we conclude that there is nothing to recognize the existence of black hole
and
therefore nothing to radiate in their cosmological model.

\vspace*{2cm}
\footnoterule
\vspace*{4mm}
\noindent
$^{(a)}$ E-mail address: tfjiwa@tansei.cc.u-tokyo.ac.jp  \\
$^{(b)}$ E-mail address: igarashi@ed.niigata-u.ac.jp\\
$^{(c)}$ E-mail address: jik@hep.s.kanazawa-u.ac.jp
\newpage
\pagestyle{plain}
\noindent
{\bf 1}  Inspired by string theory \cite{st}, much attention has
recently paid to cosmology in two-dimensions \cite{mann}. In particular,
Callan, Giddings, Harvey and Strominger (CGHS) \cite{cghs} have proposed a
two-dimensional (2D) model in which
the classically exact solutions
to field equations correspond to black holes. Their model therefore
has been considered as a theoretical laboratory to study the essence of
black hole physics \footnote{See refs. \cite{H,HS} for instance.}.

The mass parameter in their black hole
solution turns out to be arbitrary, and can not
be related to the parameters of the theory. So, we are lead to
wonder whether there is some sort of degeneracy.
One of our findings in this letter is that in fact the global symmetry
found in ref. \cite{su}
is responsible for `` degeneracy '' of the black hole back ground
and at the same time
for decoupling of the matter.
To see this, we start by writing  down
the classical action of CGHS \cite{cghs} \be
S &=& \int d^{2}\sigma\ \sqrt{-g}\,\{\,\exp (-2\phi)\,[\,
R(g)+4\,g^{\alpha\beta}\partial_{\alpha}\phi
\,\partial_{\beta}\phi +4\,\mu^2\,]\nn\\
& &-\frac{1}{2}\sum_{i=1}^{n} g^{\alpha\beta}
\ \partial_{\alpha}f_{i}
\,\partial_{\beta}f_{i}\,\}\ ,\ (~\alpha,\beta~=~0,1~)\ ,
\ee
where $f's$ stand for matter fields, $\phi$ for the dilaton field,
$\mu^2$ for a cosmological constant, and the curvature scalar
for the metric $g_{\alpha\beta}$ is denoted by $R(g)$.
In addition to 2D general coordinate transformations, the action is invariant
under the global, non-linear transformation defined by \cite{su}
\be
\phi&\rightarrow& \phi '=
\phi-\frac{1}{2}\ln (1+\Lambda\exp(2\phi)\,)~,~
g_{\alpha\beta} ~\rightarrow~g_{\alpha\beta} '\,=\,
g_{\alpha\beta}(1+\Lambda\exp (2\phi) )^{-1}~,
\ee
where $\Lambda$ is a constant parameter. The conservation of the corresponding
current, $\partial_{\alpha}j^{\alpha}=\sqrt{-\hat{g}}\,R(\hat{g})=0$,
expresses the flatness of $\hat{g}_{\alpha\beta}$,
where $\hat{g}_{\alpha\beta}\equiv g_{\alpha\beta}\exp (-2\phi)$.
(The importance
of this current conservation in investigating the Hawking radiation
in two dimensions has been emphasized in ref.
\cite{ver}, too.)

This global symmetry manifests itself in the arbitrariness of classical
solutions.
An easy way to see this is to work in the conformal gauge,
$g_{\alpha\beta}=\eta_{\alpha\beta}\exp (2\rho)$. One finds
the equations of motion,
$\partial_{+}\partial_{-}\,\exp (-2\phi)+\mu^2\exp 2(\rho-\phi) =0~,~
\partial_{+}\partial_{-}(\rho-\phi) = 0$,
along with two constraints (that are equivalent to $\varphi_{\pm}$ (8) below
in the conformal gauge).
CGHS have found that the above set of equations admits a classical solution
\be
\frac{1}{2}\,\sum_{i=1}^{n}\,(\partial_{+}f_{i})^2 &=&
-a\,\delta(\sigma^+-\sigma^{+}_{0}) ~,~ \rho=\phi=\rho_{\rm BH}~,\nn\\
\exp (-2\rho_{\rm BH}) &=& a(\sigma^+-\sigma^{+}_{0})\,\Theta
(\sigma^+-\sigma^{+}_{0})- \mu^2 \sigma^+\sigma^-~,
\ee
which describes the formation of a black hole
of mass $\mu a \sigma^{+}_{0}$ by an $f$ shock wave traveling
in the $\sigma^{-}$ -direction. However, the solution
(3) is not a unique solution for the given shock wave.  One easily
verifies that
$\exp (-2\rho)=\exp (-2\phi)= \Lambda+\exp (-2\rho_{\rm BH})$
is a solution ( with mass $\mu (a \sigma^{+}_{0}+\Lambda)$ )
too, where $\Lambda$ is an arbitrary constant.
Since under the global transformation, $\exp (-2\phi)$ and $\exp (-2\rho)$
change to $\Lambda+\exp (-2\phi) $ and $\Lambda
\exp 2(\phi-\rho)+\exp (-2\rho) $, respectively,
it is clear that
the arbitrariness of the black hole back ground originates from the
global symmetry defined by (2).

The gravitational radius in the Schwarzschild solution in $D=4$
is arbitrary, too. It is in fact possible to find
a similar symmetry there. Although this arbitrariness and the existence
of a corresponding symmetry in both dimensions are closely
related to the arbitrariness in choosing boundary conditions,
there is a
crucial difference in both dimensions:
Thanks to the Weyl invariance of the matter coupling
in $D=2$, the symmetry (2) is exact at the classical level even in the
presence of the matter.  That is,
the matter does not feel the gravity mediated by the Weyl degree of
freedom at the classical level,  unless the matter coupling violates
the Weyl invariance. (A test particle,
for instance, would feel the black hole because its coupling violates the
symmetry
(2).)

At the quantum level, Weyl symmetry is anomalous in general \cite{PolA}.
Therefore, the global symmetry (2) may be anomalous too, because it is closely
related to the Weyl invariance. But
the question of whether or not
the global symmetry really remains intact in quantum theory certainly requires
an independent investigation.

The subsequent sections  are devoted to perform an exhaustive,
algebraic analysis to find possible anomalies in the CGHS theory (1).
We will use
the method which we have developed in ref. \cite{FIKA}
on the basis of the extended phase space method of
Batalin, Fradkin and Vikovisky (BFV) \cite{FVA}.
The advantage of using this method is that the results so obtained
are independent of the choice of gauge and regularization. This feature
of the method is desirable, especially in a situation in which
anomaly might cause ``measurable'' effects, such as the Hawking radiation in
the
present case \footnote{There is in fact some gauge-dependent
observation in 2D cosmology \cite{tr}.}. Throughout this paper we shall assume
that space time
singularities do not influence the invariance property such that we may perform
our analysis without taking into account the presence of a black hole
back ground explicitly.

We shall deal with classical, cohomological problems whose cohomologically
non-trivial solutions exhibit potential anomalies. Under a certain set of
reasonable assumptions,
we will show that the global symmetry (2) as well as the BRST symmetry in the
CGHS theory are free from an anomaly, to all orders in $\hbar$.
Therefore, the matter in their model does not
feel the existence of a black hole at all, and
there is nothing to radiate \footnote{Our observation might be related to
the recent result \cite{eguchi} on the equivalence
between  $c=1$ conformal field theory and
the Wess-Zumino-Witten model based on $SL(2,R)/U(1)$.}.
\vspace{5mm}

\noindent
{\bf 2}  To apply the
BFV quantization method \cite{FVA}, we have to go to the Hamiltonian
formulation
of the theory, and we use the parametrization of $g_{\alpha\beta}$, according
to
Arnowitt, Deser, and Misner: \be
g_{\alpha\beta}&\equiv&
\left( \begin{array}{cc} -\lambda^{+}\lambda^{-} & (\lambda^{+}
-\lambda^{-})/2 \\
(\lambda^{+}-\lambda^{-})/2 & 1 \end{array} \right)\exp 2\rho ~.
\ee
In terms of these new variables, the original action can be re-written
as (we use the abbreviations $ \dot{f} = \partial_{0} f
=\partial_{\tau} f~, ~
f^{\prime} = {\partial}_{1}f ={\partial}_{\sigma}f$)
\be
S &=&\int d^2\sigma\,\{~\frac{\psi '}{\lambda^+ +\lambda^-}[\,
2\,(\lambda^+ -\lambda^-)(\dot{\rho}-\dot{\phi})
+4\,\lambda^+\lambda^-(\rho '-\phi ')
+2\,(\lambda^+ \lambda^-) '\,]\nn\\
& &+\frac{\dot{\psi}}{\lambda^+ +\lambda^-}[\,
-4\,(\dot{\rho}-\dot{\phi})
+2\,(\lambda^+ -\lambda^-)(\rho '-\phi ')
+2\,(\lambda^+ -\lambda^-) '\,]\nn\\
& &+2\mu^2\,(\lambda^+ +\lambda^-)\exp 2(\rho-\phi)\nn\\
& &+\frac{1}{\lambda^+ +\lambda^-}\sum_{i=1}^{n}\,[\,
\dot{f}_{i}\dot{f}_{i}
-(\lambda^+ -\lambda^-)\dot{f}_{i}f_{i} '
-(\lambda^+ \lambda^-)f_{i} 'f_{i} ' \,]~\}~,
\ee
where $\psi = \exp(-2\phi)$.
The conjugate momenta to $\lambda^{\pm},\rho,\phi$ and $f_{i}$
are  respectively defined as
\be
\pi_{+}^{\lambda} &=&0~,~\pi_{-}^{\lambda}=0~,\\
\pi_{\rho} &=& (\lambda^+ +\lambda^-)^{-1}\,[\,
2\,(\lambda^+ -\lambda^-)\psi '-4\dot{\psi}\,]~,\nn\\
\pi_{\phi} &=& -\pi_{\rho}-2\,(\lambda^+ +\lambda^-)^{-1}\psi\,[\,
-4\,(\dot{\rho}-\dot{\phi})+2\,(\lambda^+ -\lambda^-)(\rho '-\phi ')
+2\,(\lambda^+ -\lambda^-) '\,]~,\\
\pi_{f}^{i} &=& (\lambda^+ +\lambda^-)^{-1}\,[\,2\,\dot{f}_{i}-
 (\lambda^+ -\lambda^-)f_{i}'\,]~.\nn
\ee
As can be seen from (6), $\pi_{\pm}^{\lambda}$  are primary
constraints, and generate the secondary constraints
\be
\varphi_{\pm} &=& -2\mu^2\exp 2(\rho-\phi)
+(\partial_{\sigma}-Y_{\pm})\,(\psi '\mp\frac{1}{2}\,\pi_{\rho})
+\frac{1}{4}\,\sum_{i=1}^{n}\,(\,\pi_{f}^{i}\pm f_{i} '\,)^2~,\\
\mbox{with}  & & Y_{\pm}\equiv (\rho-\phi) '
\pm\frac{1}{4\psi}\,(\pi_{\rho}+\pi_{\phi})~,\nn
\ee
which satisfy under Poisson bracket the Virasoro algebra:
\be
\{ \varphi_{\pm} ( \sigma )\ ,\ \varphi_{\pm} ( \sigma') \}_{\rm PB} &=& \mp
(\ \varphi_{\pm} ( \sigma ) \partial_{\sigma '} - \varphi_{\pm} ( \sigma')
\partial_{\sigma}\ )\  \delta ( \sigma - \sigma')~, \\
\{ \varphi_{\pm} ( \sigma )\ ,\ \varphi_{\mp} ( \sigma')
\}_{\rm PB} &=& 0 \nn\ .
\ee
Since there are four first-class constraints, the theory without
the matter would have no physical degree of freedom
 \cite{ja}.

According to BFV, we define
the extended phase space by
including  to the classical phase space the
ghost-auxiliary field sector
\be
( {\cal C}^{\rm A}\ ,\ \overline{{\cal P}}_{\rm A} )\ ,\
( {\cal P}^{\rm A}\ ,\ \overline{{\cal C}}_
{\rm A} )\ ,\
( N^{\rm A}\ ,\ B_{\rm A} )\ ,
\ee
where  ${\rm A} ( = {\lambda}^{\pm},~\pm)$ labels the  first-class
constraints given in (6) and (8) \footnote{
$ {\cal C}^{\rm A}$ and ${\cal P}^{\rm A}$ are the BFV ghost fields
carrying one unite of ghost number,
$\mbox{gh}({\cal C}^{\rm A}) = \mbox{gh}({\cal P}^{\rm A}) =1$,
while $~~\mbox{gh}(\overline{{\cal P}}_{\rm A}) =
\mbox{gh}(\overline{{\cal C}}_ {\rm A}) = -1$
for their canonical momenta, $\overline{{\cal P}}_{\rm A}$ and
$\overline{{\cal C}}_ {\rm A}$.
The last canonical pairs in (10) are
auxiliary fields and carry no ghost number.}.
We assign $0$ to the canonical dimension of $
\phi, {\lambda}^{\pm}$ and $\rho$,
and correspondingly $+1$ to $\pi_{\phi},\pi^{\lambda}_{\pm}$, and
$\pi_{\rho}$. The canonical
dimensions of  ${\cal C}_{\lambda}^{\pm},~
\overline{{\cal P}}_{\pm}$
and $~\overline{{\cal P}}^{\lambda}_{\pm}$
are fixed only relative to that of ${\cal C}^{\pm},
 c~ \equiv \mbox{dim}({\cal C}^{\pm})$:
\be
\mbox{dim}({\cal C}_{\lambda}^{\pm})~=~1+c~ ,~
\mbox{dim}(\overline{{\cal P}}_{\pm})~ =~ 1-c~
,~  \mbox{dim}(\overline{{\cal P}}^{\lambda}_{\pm})~ =~ -c~.
\ee
The canonical
dimensions of other fields are not
needed for our purpose.

Given the first-class constraints with the corresponding algebra, we can
construct a BRST charge \be
 Q &=& \int d \sigma [\ {\cal C}_{\lambda}^{+} {\pi}^{\lambda}_{+}
+ {\cal C}_{\lambda}^{-} {\pi}^{\lambda}_{-}
 + {\cal C}^{+} ( \varphi_{+} + \overline{{\cal P}}_{+}
\partial_{\sigma} {\cal C}^{+} ) \nn \\
 & & + {\cal C}^{-} ( \varphi_{-} -\overline{{\cal P}}_{-}
\partial_{\sigma}  {\cal C}^{-} )
 + B_{\rm A} {\cal P}^{\rm A}\ ]~,\ (A=\lambda^{\pm},\pm)\\
& &\mbox{with gh}(Q) = 1 ,\mbox{dim}(Q)=1+c~,\nn
\ee
which satisfies
\be
\{Q~,~Q\}_{\rm PB} &=&0~.
\ee
We would like to emphasize that the BRST charge (12) is given prior to gauge
fixing. The gauge fixing appears in defining the total Hamiltonian $~H_{\rm
T}$.
Since the canonical Hamiltonian
vanishes in the present case,
it is given by the BRST variation of gauge fermion ${\Psi}$,
$H_{\rm T}=\{ Q~,~\Psi \}_{\rm PB}$,
which immediately leads to
\be
\{Q\ ,\ H_{\rm T}\}_{\rm PB}\ &=&\frac{d}{dt}Q~ =~ 0\ .
\ee

In terms of the phase space variables,
the charge $W$, which generates the non-linear symmetry transformation (2),
can be written as
\be
W &=&-\frac{1}{2}
\int d\sigma\,(\pi_{\phi}+\pi_{\rho})\psi^{-1}~=~
-\int d\sigma\,(Y_{+}-Y_{-})~,
\ee
where $Y_{\pm}$ are defined in (8), and its
Poisson brackets with $Q$ and $H_{\rm T}$ are
\be
\{Q~,~W \}_{\rm PB} &=&0~, \\
\{H_{\rm T},~W\}_{\rm PB} &=&0~.
\ee
Eqs. (13), (14), (16) and (17) are the basic bracket relations which
exhibit the  2D general covariance and the global invariance at the classical
level.
\vspace{0.5cm}

\noindent
{\bf 3}  The ``nilpotency of $Q$" expressed in eq. (13) means that the
underlying  constraints in the theory are first-class while eq. (14)
expresses  the consistency of the constraints with the dynamics of the
system. Eqs. (16) and (17) exhibit the presence of the global
symmetry. Note however that because of the relation,
$H_{\rm T}=\{ Q~,~\Psi \}_{\rm PB}$,
eqs. (14) and (17)
are consequences of (13) and (16). We therefore shall
consider only
(13) and (16) as the fundamental bracket relations .

At the quantum level, these quantities must be suitably regularized
to become well-defined.
An anomaly arises if
the fundamental algebra between $Q$ and $W$ can not be maintained upon
quantization, and the anomalous Schwinger terms, which exhibit an anomaly in
the
algebra, may be expanded in $\hbar$ as
\be [Q\ ,\ Q] &\equiv & i\sum_{n=2}\hbar^n\Omega^{(n)}~,~
[ Q\ ,\ W] ~\equiv~ {i\over2}\sum_{n=2}\hbar^n \Xi^{(n)}\ ,
\ee
where $[\ ,\ ]$ denotes super-commutator.
Our basic assumption \cite{FIKA,FIKM} is that the super-commutation relations
between $Q$ and $W$ obey :
(i) the commutation law, (ii)
the distribution law, (iii) the super-Jacobi identity, and
(iv) $[A~,~B]~=~i\hbar\,\{A~,~B\}_{\rm PB}+O(\hbar^2)$.
One easily observes that the outer commutators in the super-Jacobi
identities for $Q$ and $W$,
\be
[Q\ ,[Q\ ,\ Q]\ ]&=&0\ ,~
 2\ [Q\ ,\ [Q\ ,\ W]\ ]+[W\ ,\ [Q\ ,\ Q]\ ]=0\ ,
\ee
define a set of consistency conditions at each order in $\hbar$ in terms of
Poisson brackets.
In the lowest order, one finds that \cite{FIKA,FIKM}
\be
\delta\Omega^{(2)} &= &0~,\\
\delta\Xi^{(2)} &= &\{W\ ,\ \Omega^{(2)}\}_{\rm PB}\  ,
\ee
where $\delta$ is the BRST transformation which is defined by
$\delta A = - \{Q\ ,\ A\}_{\rm PB}$.
The true anomalies $\Omega^{(2)}$ and $\Xi^{(2)}$ should be
cohomologically non-trivial;
if $\Omega^{(2)}$ and $\Xi^{(2)}$ are solutions, then $\Omega^{(2)} + \delta X$
and
$\Xi^{(2)} + \{ W\ ,\  X\}_{\rm PB} + \delta Y$
also solve (20) and (21), respectively. They can be then absorbed to order
$\hbar^2$ into a re-definition of $Q$ and $W$, defined by
$Q\rightarrow  Q-(\hbar X/2)~\mbox{and}~  W\rightarrow
W - (\hbar Y/2)$.
Furthermore, if there is no non-trivial solution to (20)
(this is the case in the CGHS theory as we shall see later),
the non-trivial $\Xi^{(2)}$ is a non-trivial solution to the homogeneous
part of (21):
\be
\delta\,\Xi^{(2)}_{h}&=&0~.
\ee

At this stage we would like to emphasize that the consistency conditions
of $O(\hbar^2)$ are exactly the same as those of $O(\hbar^3)$ if
there is no non-trivial solution at $O(\hbar^2)$. That is, the non-existence
of the non-trivial solutions to the consistency conditions (20) and (22)
is sufficient for the non-existence
of those to the higher order consistency conditions.

Our main task is to solve the classical, algebraic problem defined
by (20) and (21) (or (22) if there is no non-trivial solution to (20)).
We seek solutions $\Omega^{(2)}$ and $\Xi^{(2)}$ in the form
\be
\Omega^{(2)} &=& \int d \sigma \omega\ , ~
\Xi^{(2)} ~(\Xi^{(2)}_{h})~=~ \int d \sigma \xi~(\xi_{h})\ ,
\ee
where $\omega$  and $\xi~ (\xi_{h})$ are polynomials of local operators with
$~\mbox{gh}(\omega) = 2~$, $~\mbox{gh}(\xi) = \mbox{gh}(\xi_{h})=1~$,
$~\mbox{dim}(\omega) = 3~+~2c$, and $~\mbox{dim}(\xi) = \mbox{dim}(\xi_{h})
=2~+~c$.  According to the general structure of the BFV formalism, the total
phase space can be divided, with respect to the action of $\delta$,
into two sectors;
\be
& &{\rm S}_{1}\ \mbox{ consisting of}\
(f_{i}\ ,\ \pi_{f}^{i})\ ,(\rho\ ,\ \pi_{\rho})\ ,\ (\phi\ ,\ \pi_{\phi})\
 \mbox{and}\ ({\cal C}^{\pm}\ ,\ \overline{{\cal P}}_{\pm})\ ,  \nn\\
& &
{\rm S}_{2}\ \mbox{consisting of all the other fields}\ .
\ee
It is easy to see that on each sector the $\delta$ operation closes:
${\delta_{1}}^2 = {\delta_{2}}^2\ =\ 0\ ,\
\delta_{1} \delta_{2} + \delta_{2} \delta_{1}\  =\ 0$,
where $\delta =\delta_{1} + \delta_{2}$, and $\delta_{1} (\delta_{2})$ acts on
${\rm S}_{1}$ (${\rm S}_{2}$) variables only.
The ${\rm S}_{2}$-sector is BRST trivial because it is made of pairs
$(U^{a}\ ,\ V^{a})$
 with $\delta_{2} U^{a}= \pm V^{a}$.
As shown in ref. \cite{FIKA}, there exists no non-trivial solution to (20)
((22)) if $\omega\,(\xi_{h})$  contains the ${\rm S}_{2}$-variables.

The linear independence of the generalized Virasoro constraints
\be
\Phi_{\pm}&\equiv&-\delta\,\overline{{\cal P}}_{\pm}~ = ~\varphi_{\pm} \pm
[ \, 2 \overline{{\cal P}}_{\pm} \, {\cal C}^{\pm '}+ \overline{{\cal
P}}_{\pm}'
\, {\cal C}^{\pm} \, ]~
\ee
and the fact that $\overline{{\cal P}}_{\pm}$ is the only one which produces
$\overline{{\cal P}}_{\pm}$  under BRST transformation further implies that
$\overline{{\cal P}}_{\pm}$ can not be involved in the non-trivial part
of $\omega$ and $\xi_{h}$.
Therefore, $\omega$ and $\xi_{h}$ are functions
of $f_{i}, \pi_{f}^{i},\rho, \pi_{\rho} ,\phi , \pi_{\phi}$,
${\cal C}^{\pm}$ and their spatial derivatives
only \footnote{We regard spatial total derivative terms as null.}.

To proceed, we
note that there is a time-independent reparametrization invariance
(i.e., $\{\Phi_{\pm}~,~Q ~\mbox{and}~W\}_{\rm PB}=0$ ).
Its transformation can be generated by $\Phi_{\pm}$, and
it is convenient to define
covariant objects with respect to the transformations.
The $Y_{\pm}$ given in eq.
(8), for instance, are `` gauge fields ''  transforming as
\be
\delta_{\pm}\,Y_{\pm} &=& \pm u^{\pm}(\sigma)''\pm[\,u^{\pm}
(\sigma)\,Y_{\pm}\,]'~,
\ee
where
$
\delta_{\pm} \,\cdot\equiv -\{\Phi_{\pm}^{u}~,~\cdot\}_{\rm PB}
{}~\mbox{with}~\Phi_{\pm}^{u}\equiv \int d\sigma\,
u^{\pm}(\sigma)\,\Phi_{\pm}$.
Then we may define
the weight $w_{\pm}$ of a field $\chi$, according  to
\be
\delta_{\pm} \chi &\equiv& \pm u^{\pm}\,\chi ' \pm w_{\pm}\, u^{\pm '} \chi~,
\ee
and the covariant derivatives
by
\be
D_{\pm} &\equiv& \partial_{\sigma}-Y_{\pm}\,w_{\pm}~.
\ee
Note that $\rho$ and $\phi$ do not transform covariantly
and there is no bosonic field of weight zero. The only covariant
quantities are \footnote{We have dropped
$~\overline{{\cal P}}_{\pm}
{}~(\mbox{with}~w_{\pm}=2) $ from the list because it is not
involved in $\omega$ and $\xi~(\xi_{h})$.}
\be
Y_{\pm} & & \mbox{gauge fields}~,\nn\\
{\cal C}^{\pm} & & \mbox{with}~w_{\pm}=-1~,\nn\\
F_{\pm}^{i}&\equiv &\pi_{f}^{i}\pm f_{i}'~\mbox{with}~~~w_{\pm}=1~,\\
G_{\pm}&\equiv  &-2\mu^2\exp 2(\rho-\phi)+
(\partial_{\sigma}-Y_{\pm})\,(\psi
'\mp\frac{1}{2}\,\pi_{\rho})
 ~\mbox{with}~w_{\pm}=2~,\nn
\ee
and those which can be obtained by successive applications
of the covariant derivatives thereon.
 The BRST transformations of these quantities
are closed,
as one can see from
\be
\delta\,Y_{\pm} &=& \pm(\,D_{\pm}{\cal C}^{\pm}\,)'~,~\delta\,{\cal C}^{\pm}
=\pm{\cal C}^{\pm}{\cal C}^{\pm '}~,\nn\\
\delta\,F_{\pm}^{i} &=&\pm(\,{\cal C}^{\pm}\,F_{\pm}^{i}\,) ' ~,~
\delta\,G_{\pm} =\pm {\cal C}^{\pm}\,G_{\pm} '
\pm 2  {\cal C}^{\pm '}\,G_{\pm}~.
\ee

At this stage, we make our main assumption to reduce the complexity
of our cohomological problem: We assume that
the non-trivial parts of $\omega$ and $\xi_{h}$
are functions only of the quantities in (29) and their
spatial derivatives  and that they respect the discrete symmetry defined by
${\cal C}^{\pm} \rightarrow {\cal C}^{\mp},\overline{{\cal P}}_{\pm}
\rightarrow \overline{{\cal P}}_{\mp},
\partial_{\sigma} \rightarrow - \partial_{\sigma}$.

With this assumption in mind,
we then group all the possible terms
that may be
present in $\omega$ and $ \xi_{h}$:
\be
& &{\rm g}_{1(2)}\ \mbox{ consisting of terms containing at least one}\
G_{\pm}~(F_{\pm}^{i})
{}~,\\
& & {\rm g}_{3}\ \mbox{ stands for the rest}\ .\nn
\ee
It is clear that BRST transformations do not mix the terms
of different groups.

We begin to solve (20). Since $\omega~(\mbox{dim}(\omega)=3+2c)$
must contain two ${\cal C}$'s,
each term of $g_{1}$ for $\omega$ has just one of
$G_{\pm}~(\mbox{dim}(G_{\pm})=2)$. Exactly
four terms that contain besides one of the gauge fields $Y_{\pm}$
come into the question. One can easily verify that
they can never organize to a BRST invariant. As for the rest of $g_{1}$, one
can
write down four independent terms for $\omega$ that are consistent with our
assumption. We find that all the
BRST invariants are trivial, namely proportional to $\delta\,
[\,(\kappa_{1}\,{\cal C}^{+}\,G_{+}- \kappa_{2}\,{\cal
C}^{-}\,G_{+})-(+\leftrightarrow -)\,]$, where $\kappa_{1,2}$ are
arbitrary constants.
The $g_2$-elements for $\omega$ can be further divided into two groups; the one
consisting of terms with one of $\sum_{i=1}^n\,F_{\pm}^{i} F_{\pm}^{i}$, and
the
other one consisting of terms with one $\sum_{i=1}^n\,F_{+}^{i} F_{-}^{i}$.
The BRST-transformation property of $\sum_{i=1}^n\,F_{\pm}^{i} F_{\pm}^{i}$
is the same as that of $G_{\pm}$, and this is the reason why
the terms of the first group are absent in $\omega$. However, the
second group of $g_{2}$ contains one non-trivial BRST invariant \cite{FIKM}:
\be
\omega_{2} &=& \kappa_{3}\,[~({\cal C}^{+}
\,{\cal C}^{+ '}+{\cal C}^{-}{\cal C}^{+ '})-(+\leftrightarrow -)~
]\,\sum_{i=1}^n\,F_{+}^{i}F_{-}^{i}~.
\ee
This matter-field dependent term
$\omega_{2}$ is algebraically
allowed as a $Q^2$-anomaly, but such
matter-field dependent expressions have never appeared in the explicit
calculations of anomalies. This may easily be understood if one employs
Fujikawa's method \cite{kf} to calculate anomalous terms \footnote{See ref.
\cite{IKS} for
a method
to calculate anomalous commutators from anomalous path-integral Jacobians.}.
Therefore, we demand
\be
\kappa_{3} &=&0~.
\ee

The third group, $g_{3}$, contains
the Kato-Ogawa anomaly term \cite{KOA},
\be
\omega_{\rm KO} &=& \kappa_{\rm KO}\,[\,({\cal C}^{+ '}\,
{\cal C}^{+''})-( +\leftrightarrow -)\,]~,
\ee
which however
 is trivial  because of the identity
\be
{\cal C}^{\pm '}\, {\cal C}^{\pm ''} &=&\mp\delta
\,({\cal C}^{\pm '}\,Y_{\pm}) +({\cal C}^{\pm}{\cal C}^{\pm '}Y_{\pm}) '\ .
\ee
After similar algebraic calculations, one finds that no non-trivial
BRST invariant in $g_{3}$ can be formed to become an independent part of
$\omega$.

We finally would like to come to the solution of (21). Since $Q^2$ has turned
out to be trivial, the true anomaly for the global non-linear symmetry
corresponds to the non-trivial solution to the homogeneous equation (22).
We have to perform algebraic calculations, similar to the previous ones
but with ghost number and canonical dimension changed
($\mbox{gh}(\xi_{h})=1~,~\mbox{dim}(\xi_{h})=2+c$).
For $\xi_{h}$ there are two independent terms in $g_{1}$, consistent with
our assumption. It is easy to find that they can not give any BRST invariant.
Similarly, terms in $g_{2}$ with $\sum_{i=1}^n\,F_{\pm}^{i} F_{\pm}^{i}$
can not be present in $\xi_{h}$. But, as the case
for $\omega_{2}$, there is exactly one BRST invariant in $g_{2}$:
\be
\xi_{2} &=& \nu_{3}\,({\cal C}^{+}
+{\cal C}^{-})\,\sum_{i=1}^n\,F_{+}^{i}F_{-}^{i}~,
\ee
where $\nu_{3}$ is an arbitrary constant.
Terms in $g_{3}$ for $\xi_{h}$ must contain at least one of $Y_{\pm}$.
Those with $(Y_{+})^2$ or $(Y_{-})^2$ can be simply excluded.
The only possibility is
\be
\xi_{3} &=& \nu_{4}\,[\,D_{+}{\cal C}^{+}\,Y_{-}+D_{-}{\cal C}^{-}\,Y_{+}\,]~,
\nn\\
&=&\nu_{4}\,\delta [(\rho-\phi)\,(Y_{-}-Y_{+})\,]
-\nu_{4} \,\{\,(\rho-\phi)(D_{+}{\cal C}^{+}+D_{-}{\cal C}^{-})\,\}' ~,
\ee
which is trivial as the second equation indicates.

The same reason why $\omega_2 =0 $ (see (33)) can be applied to exclude
that matter-field dependent term $\xi_{2}$ as an independent anomalous
Schwinger term. We thus arrive at the conclusion that there is no non-trivial
solutions to (20) and (22) and hence the BRST symmetry and
the global symmetry (2) are exact to all orders in $\hbar$.

The general result on the trace anomaly of ref. \cite{PolA}
and the analysis on
the relation between the trace anomaly and  Hawking radiation
\cite{C} remain of course
correct. But what we have found here implies that due to the very nature
of the dilaton field the trace anomaly term can be absorbed into a
re-definition of the various fields without violating the
reparametrization invariance.

\vspace{0.5cm}

\noindent
{\bf 4}  We thus have shown that the CGHS theory is free from the BRST anomaly
and
has an exact global invariance (2) that is responsible for decoupling
of the matter. It is certainly worthwhile to collect our assumptions which have
led to the conclusion: We have assumed that (I) the black hole background does
not
influence the invariance property of the theory so that we may perform
our analysis on anomalies without taking into account its presence explicitly,
(II) the commutators of $Q$ and $W$ satisfy the super-Jacobi identities (19),
(III) the anomalous commutators can be expanded in $\hbar$, and
(IV) the non-trivial solutions to the consistency conditions (20) and (22)
involve only the quantities listed in (29).

The main reason why the BRST invariance in the CGHS model is intact
is that
the Kato-Ogawa anomaly term (34)
(that corresponds to the central extension of the Virasoro
algebra) is BRST trivial. That
is, this term can be canceled by adding to the action a local counterterm
of the form
\be
-\frac{\hbar}{2}\kappa_{\rm KO}\int d^2 \sigma\,(\,
N^+Y_+'+N^-Y_-'\,)\ ,
\ee
which becomes
\be
& &-\frac{\hbar}{2}\kappa_{\rm KO}\int d^2 \sigma\,
\frac{1}{N^++N^-}\,[\,2(N^+-N^+)'(\dot{\rho}-\dot{\phi})\nn\\
& &+2(N^+N^-)'(\rho '-\phi ')-(N^{+'}-N^{-'})^2\,]\ ,
\ee
in the standard gauge ( $N^+=\lambda^+,
N^-=\lambda^-$) \cite{FIKM}. The integrand is a total derivative in the
conformal gauge ( $N^+=\lambda^+=N^-=\lambda^-=1$), as one can easily
see from (39).  The contribution of this counterterm to the energy
momentum tensor in the conformal gauge is
\be
\Delta T_{\pm\pm} &=&\frac{\hbar}{2}\kappa_{\rm KO}
Y_{\pm} '=  \hbar\,\kappa_{\rm KO}\,\partial_{\pm}^2(\rho-\phi)
\ ,\ \Delta T_{+-}\,=\,0\ ,
\ee
which vanish on the black hole back ground (3).
This should be contrasted to the case in string theory
where any counterterm to cancel the term (34) necessarily contributes
to the trace of the energy momentum tensor
in a non-trivial manner \cite{PolA}.

An independent verification of our result, that
the BRST symmetry is anomaly-free and it is possible to
redefine the charge $W$  so as to be conserved and BRST invariant,
by an explicit computation would be of course desirable.
(There is some indication \cite{tr,hamada}.) In this context it may be
worth-mentioning that
the case at hand is quite similar to that of the
ghost-number current anomaly in string theory \cite{FujB} because
at subcritical dimensions it is possible  to redefine the charge
associated with the ghost number conservation so as to commute with
the BRST charge as well as the Hamiltonian \cite{FujB}.
\vspace{10mm}

We thank K. Fujikawa for careful reading of the manuscript and suggestions,
and K. Itoh and H. Terao for discussions.


\end{document}